\documentclass[aps,prb,twocolumn,showpacs,superscriptaddress,amssymb,amsmath,floatfix]{revtex4}
\usepackage{graphicx}

\usepackage{amsfonts}
\begin{document}

\title{\bf Phonon Hall effect in ionic crystals in the presence of static magnetic field}
\author{Bijay K. Agarwalla}
\affiliation{Department of Physics and Centre for Computational Science and Engineering, National University of Singapore, Singapore, 117542, Republic of Singapore}
\author{Lifa Zhang}
\affiliation{Department of Physics and Centre for Computational Science and Engineering, National University of Singapore, Singapore, 117542, Republic of Singapore}
\author{Jian-Sheng Wang}
\affiliation{Department of Physics and Centre for Computational Science and Engineering, National University of Singapore, Singapore, 117542, Republic of Singapore}
\author{Baowen Li}
\affiliation{Department of Physics and Centre for Computational Science and Engineering, National University of Singapore, Singapore, 117542, Republic of Singapore}
\affiliation{NUS Graduate School for Integrative Sciences and Engineering, Singapore 117456, Republic of Singapore}


\begin{abstract}
  We study phonon Hall effect (PHE) for ionic crystals in the presence of static magnetic field. Using Green-Kubo formula, we present an exact calculation of thermal conductivity tensor by considering both positive and negative frequency phonons. Numerical results are shown for some lattices such as hexagonal lattices, triangular lattices, and square lattices. We find that the PHE occurs on the nonmagnetic ionic crystal NaCl, although the magnitude is very small which is due to the tiny charge-to-mass ratio of the ions. The off-diagonal thermal conductivity is finite for nonzero magnetic field and changes sign for high value of magnetic field at high temperature. We also found that the off-diagonal thermal conductivity diverges as $\pm{1/T}$ at low temperature.
\end{abstract}


\maketitle

\section{INTRODUCTION}

The magnetic field dependence of heat conductivity in paramagnetic dielectric crystals has been found experimentally \cite{strohm05}. This effect is known as ``phonon Hall effect''. Applying magnetic field in the direction normal to the temperature gradient, the authors have discovered the appearance of a transverse thermal current in Tb$_3$Ga$_5$O$_{12}$ and the effect was confirmed in Ref.~\onlinecite{inyushkin07}. Due to the Lorentz force on the electrons, electronic Hall effect can be easily understood. However for phonons there is no such direct coupling with the magnetic field. The underlying mechanism that determines this effect is the spin-phonon interaction (SPI)\cite{old-sp-papers,sp-book,ioselevich95} of phonons and paramagnetic ions Tb$^{+3}$. This effect is an analog of the anomalous Hall effect (AHE) in paramagnetic phase. The theory of this phenomenon was proposed in Refs.~\onlinecite{sheng06,kagan08,wang09}, where the authors consider the spin-phonon interaction and solve the problem perturbatively. The same effect was also studied in four-terminal junctions using nonequilibrium Green's function (NEGF) approach \cite{lifa09}. In Ref.~\onlinecite{maksimov} the authors explained that molecules with rotary degree of freedom can also have the same effect as PHE through renormalization of acoustic waves. This effect for gases is knows as Senftleben-Beenakker effect. Very recently, the topological nature of PHE and an associated phase transition was found in the paramagnetic dielectrics \cite{zhang10}. 

All the previous work on PHE \cite{strohm05,sheng06,kagan08,wang09,lifa09,zhang10} are based on the paramagnetic dielectrics; and the recent work \cite{onose10} about the magnon Hall effect in the ferromagnetic insulator could also include the PHE to the thermal Hall effect. We know that all the materials in the previous experimental or theoretical work on PHE are paramagnetic or ferromagnetic ionic crystals; therefore, a question arises promptly: whether  the PHE can occur in a general nonmagnetic ionic crystal? From a theoretical point of view, it is highly desirable to study both the existence and properties of the PHE in general ionic crystals with an applied magnetic field, which could also guide us to do the further experimental study on PHE.

As there is no magnetization effect in the nonmagnetic ionic crystal, intuitively one can not imagine the existence of PHE in the sample. The Hamiltonian in the harmonic approximation for the ionic crystal in the presence of a magnetic field is quadratic and is very similar to the form in Refs.~\onlinecite{sheng06,kagan08,wang09}. Because of the quadratic nature of the problem it is possible to diagonalize the Hamiltonian in terms of creation and annihilation operators. In Ref.~\onlinecite{wang09} the authors consider the Hamiltonian which is not positive definite and omit the contribution of negative frequency to the phonon Hall conductivity which was incorrect. Here we carefully reinvestigate the Hamiltonian of the ionic system, the non-positive-definite problem is naturally solved and the phonon Hall conductivity is calculated exactly by applying the Green-Kubo formula in the ballistic thermal transport region.  The paper is organized as follows. In sec.~II we first introduce the model, then outline the derivation of second quantization and also of thermal conductivity. In sec.~III we present numerical results some lattices, and also apply our method to the sodium chloride (NaCl) crystal in Sec.~IV. Finally we conclude with a short discussion in sec.~V.

\section{THE HAMILTONIAN IN A UNIFORM MAGNETIC FIELD}

The Hamiltonian operator of a crystal lattice in the presence of a magnetic field is, according to the minimal energy principle, given by \cite{Holz}
\begin{eqnarray}
H &=& \frac{1}{2} \sum_{{l,x,\alpha}}M_{x}^{-1}\bigl(P_{\alpha}(l,x)-eA_{\alpha}(l,x)\bigr)^{2} \nonumber \\
  &+& \frac{1}{2}\sum_{ll'}\sum_{xx'}\sum_{{\alpha}{\beta}}U_{\alpha}(l,x)\phi_{{\alpha}{\beta}}(l,x;l',x')U_{\beta}(l',x').
\end{eqnarray}
Here $M_{x}$ and $e$ are the mass and charge respectively of the $x$-th ion in the unit cell, which is assumed to have $r$ ions, $U_{\alpha}(lx)$ is the $\alpha$-th cartesian coordinate of the displacement of the $x$-th atom in the $l$-th unit cell from the equilibrium position $x_{\alpha}(lx)$, $P_{\alpha}(lx)$ is the corresponding linear momentum operator, $\phi_{{\alpha}{\beta}}(l,x;l',x')$ are the atomic force constants for the crystal and ${\bf A}$ is the magnetic vector potential.
In the presence of Lorentz gauge ($\nabla \cdot{\bf A} =0$) the vector potential can be written as a linear combintation of equilibrium position of each ion seperately  $A_{\alpha}(l,x)=\sum_{\beta} C_{{\alpha}{\beta}}U_{\beta}(l,x)$ where $C$ is an antisymmetric $3 \times 3$ matrix.
$P_{\alpha}(l,x)$ and $U_{\alpha}(l,x)$ satisfy the commutation relation
\begin{equation}
[U_{\alpha}(l,x),P_{\beta}(l',x')]=i \hbar \delta_{{\alpha}{\beta}} \delta_{ll'} \delta_{xx'}.
\end{equation}
Using this commutation relation and substituting the expression for ${\bf A}$, the Hamiltonian of the system can be written in a compact form
\begin{equation}\label{eq-ham}
  H = \frac{1}{2} p^T p + \frac{1}{2} u^T K u + u^T {\tilde A} p,
\end{equation}
where $u$ denotes the column vector of displacements of the atoms away from the equilibrium positions for all the degrees of freedoms, multiplied by a  corresponding square root of mass $\sqrt{M_{x}}$, $p$ is the associated conjugate momenta. Superscript $T$ stands for transpose. The $K$ matrix has the following form
\begin{equation}
K_{{\alpha x},{\beta x'}}(l,x;l',x')=\frac{{\phi_{{\alpha}{\beta}}(l,x;l',x')}}{\sqrt{M_{x}M_{x'}}}-({\tilde A})_{{\alpha x},{\beta x'}}^{2}\delta_{ll'},
\end{equation}
where the matrix elements of ${\tilde A}$ are the cyclotron frequency for different masses and is given by
\begin{equation}
{\tilde A}_{{\alpha x},{\beta x'}}=\frac{e}{M_{x}}C_{{\alpha}{\beta}}\delta_{xx'}.
\end{equation}
If we take the magnetic field ${\bf B}$ along $z$ direction, ${\bf B}=B \hat{z}$, then the typical form of ${\tilde A}$ matrix would be
\begin{equation}
\left(\begin{array}{rrr} 0 & \Omega & 0  \\
         -\Omega & 0  & 0  \\
          0 & 0 & 0 \\
       \end{array} \right).
\end{equation}
where $\Omega=$$eB$/$M$.
The term $u^T {\tilde A} p$ can be treated as onsite potential which was introduced phenomenologically in all other models to describe PHE but it comes out naturally for ionic crystals. Also because of the special form of the $K$ matrix ($K=\phi-{\tilde A}^{2}$) and ${\tilde A}$ being antisymmetric matrix the system is indeed stable for large values of magnetic field.

Using the Heisenberg equation of motion and Bloch's theorem we can write the equation of motion for $u$ and $p$ in the following matrix form
\begin{equation}
  {\cal M}  \chi = \omega \chi
\end{equation}
where
\begin{equation}
{\cal M}= -i\left( \begin{array}{cc} {\tilde A} & D \\
                      -I & {\tilde A} \end{array} \right) ,
\label{eq-eigen}
\end{equation}
 $\chi_{\sigma}({\bf k}) = (\mu_{\sigma}({\bf k}), \epsilon_{\sigma}({\bf k}))^T$, $\mu_{\sigma}({\bf k})$ and $\epsilon_{\sigma}({\bf k})$ are the polarization vectors corresponding to $p$ and $u$, respectively, $D({\bf k})=\sum_{l'} K_{ll'} e^{i {\bf k}\cdot ({\bf R}_{l'}-{\bf R}_{l})}$ is the dynamical matrix, ${\bf R}_{l}$ is the real-space lattice vector and $I$ is an identity matrix.

The displacement and momentum operators can be written in the following standard second quantization form
\begin{eqnarray}
u_l &=& \sum_{{\bf k},\sigma} \epsilon_{\sigma}({\bf k}) e^{i {\bf R}_l \cdot {\bf k}} \sqrt{\frac{\hbar}{2\left|\omega_{\sigma}({\bf k})\right | N}} \, a_{\sigma}({\bf k}); \nonumber\\
p_l &=& \sum_{{\bf k},\sigma} \mu_{\sigma}({\bf k}) e^{i {\bf R}_l \cdot {\bf k}} \sqrt{\frac{\hbar}{2\left|\omega_{\sigma}(\bf k)\right| N}} \, a_{\sigma}({\bf k}).
\end{eqnarray}
Where ${\bf k}$ is the wavevector, $N$ is the number of unit cells, $\sigma$ is the phonon branch index which runs over both positive and negative frequencies, $\left|\omega_{\sigma}({\bf k})\right |=\omega_{\sigma}({\bf k}) \rm{sign}(\sigma)$ and $a_{-\sigma}({\bf k})=a_{\sigma}^{\dagger}(-{\bf k})$. The reason that we have negative frequencies is because they are also the eigenvalues of Eq.(7). Only both the positive and negative frequency eigenmodes form a complete set. Using this complete set of eigenvalues and eigenvectors the Hamiltonian can be re-expressed as
\begin{equation}
H=\frac{1}{2}\sum_{{\bf k},\sigma} \hbar \left|\omega_{\sigma}({\bf k})\right| a_{\sigma}^\dagger({\bf k}) a_{\sigma}({\bf k}).
\end{equation}
Using the defination of local energy density and continuity equation
\begin{eqnarray}
\nabla \cdot S(x)+\frac{\partial H(x)}{\partial t}=0, \\
S=\frac{1}{V}\int d^{3}x S(x),
\end{eqnarray}
total heat current operator can be written as \cite{hardy63,sheng06,kagan08,wang09}
\begin{equation}
S^{a}=\frac{1}{2}\sum_{l,l'}(R_{l}^{a}-R_{l'}^{a})u_{l}^T K_{l,l'} \dot{u}_{l'},
\end{equation}
where $\dot{u_{l}}=p_{l}-{\tilde A}u_{l}$. The index $a$ corrensponds to different cartesian axes. $V$ is the total volume of $N$ unit cells. In terms of creation and annihilation operators the above expression can be written in an exact form as
\begin{equation}
S^{a}=\frac{\hbar}{4V}\sum_{k,\sigma,\sigma'} F_{\sigma,\sigma'}^{a}({\bf k})\frac{\omega_{\sigma'}({\bf k})}{\sqrt{\left|\omega_{\sigma}({\bf k})\right| \left|\omega_{\sigma'}({\bf k})\right|}} a_{\sigma}^\dagger({\bf k}) a_{\sigma'}({\bf k}),
\end{equation}
where the $F$ function is defined as
\begin{equation}
F_{\sigma\sigma'}^a({\bf k}) =
\epsilon_{\sigma}^\dagger({\bf k}) \frac{\partial D({\bf k})}{\partial{\bf k}^a} \epsilon_{\sigma'}({\bf k}).
\end{equation}
$F$ is a Hermitian matrix with real diagonal elements and is related to the group velocity $v_{\sigma}({\bf k})$ as $ F_{\sigma\sigma}^a({\bf k}) =2  v_{\sigma}({\bf k}){\omega}_{\sigma}({\bf k})$ Off-diagonal elements are in general not zero and is responsible for carrying heat in Hall effect. The above expression of $S_{a}$ contains all four possible combinations of creation and annihilation operators which are $a_{\sigma}^{\dagger}({\bf k})a_{\sigma'}({\bf k})$, $a_{\sigma}({\bf k})a_{\sigma'}^{\dagger}({\bf k})$, $a_{\sigma}^{\dagger}({\bf k})a_{\sigma'}^{\dagger}(-{\bf k})$ and $a_{\sigma}(-{\bf k})a_{\sigma'}({\bf k})$.
The measured heat current ${\bf J}=\rm{Tr}(\rho {\bf S})$ vanishes in equilibrium. The thermal conductivity tensor can be calculated from the Green-kubo formula \cite{mahan00,allen93}
\begin{equation}
\kappa_{ab} = \frac{V}{T} \int_0^{\beta}\!\!\!\! d\lambda
\int_0^\infty\! dt\, \bigl\langle S^a(-i\lambda \hbar) S^b(t) \bigr\rangle_{\rm eq},
\end{equation}
where $\beta=1/(k_BT)$, the average is taken over the equilibrium ensemble with Hamiltonian $H$. Substituting
the expression $S^a$ and using Wick's theorem we obtain

\begin{eqnarray}
\kappa_{ab} &=& \frac{\hbar}{32VT} \sum_{{\bf k}, \sigma \ne \sigma'}
 F_{\sigma'\sigma}^a({\bf k}) \times
        F_{\sigma\sigma'}^b({\bf k})\times \frac{(\omega+\omega')^2}{\omega\omega'}\qquad\nonumber \\
    && \times\frac{n(\omega')-n(\omega)}{\omega'-\omega} \times
 \frac{1}{\eta - i (\omega- \omega')},
\label{eq-main}
\end{eqnarray}
where $n(\omega)=(e^{\beta \hbar \omega}-1)^{-1}$ is the Bose distribution
function. We have used the notations $\omega= \omega_{\sigma}({\bf k})$, $\omega'=\omega_{\sigma'}({\bf k})$. $\eta$ comes from the damping term $e^{-\eta t}$ which we have added in order to integrate the oscillatory factor $e^{i (\omega-\omega')t}$. Same phonon branch ($\sigma=\sigma'$) does not contribute to the thermal conductivity. For high temperature ($\beta \hbar \omega<1$) thermal conductivity reaches a constant value but for low temperature ($T \rightarrow 0$) it diverges as $\pm{1/T}$. For a perfect crystal without the magnetic field($A$=0) the diagonal elements of thermal conductivity $\kappa_{aa}$ diverge in the form of $1/\eta$ , corresponds to infinite conductivity and the off-diagonal elements $\kappa_{ab}$ are zero which is consistent with  Fourier's law. For nonzero magnetic field ($A \ne 0$) the diagonal elements of $\kappa$ are still infinite but off-diagonal elements gives rise to finite value of conductivity. The form of Eq.(17) is very similar to the anomalous Hall effect for electrons where the summation is replaced by an integral of ${\bf k}$ \cite{nagosa09}. The same form was also derived for disordered harmonic crystal and for amorphous solids \cite{allen93,allen89}. The formula can also be considered from a topological point of view \cite{zhang10}. The major difference of our result with others\cite{wang09,kagan08,allen93} is in the expression for $S^{a}$ where we have taken all possible combinations of creation and annhilation operators. As a result we need to sum over all positive and negative phonon branches. We numerically calculate the conductivity and found that the effect is almost twice in comparison with Ref.~\onlinecite{wang09}.

\section{CALCULATION ON SOME LATTICES}
Before studying the real ionic-crystal material, we calculate the PHE on some lattices to get some basic properties of the PHE in the ionic lattices. In this section we show the calculation of the PHE on the hexagonal lattices, triangular lattices, and square lattices.
  
In Fig.\,1, we give results of dispersion relation for a hexagonal lattice with nearest neighbour couplings. The primitive vectors for the lattice are ($a$,0) and ($a$/2,$\sqrt{3} a$/2) with $a=1$ \AA. The longitudinal and transverse force constants are chosen as $K_{L}\,=\,0.144$ eV/(u\AA$^2)$ and $K_{T}=0.036$ eV/(u\AA$^2)$ respectively\cite{wang09}. This typical choice is comparable with experimental values. Here $h=eB/m$ which has dimension of frequency.   In the presence of magnetic field the phonon spectrum is shifted (see Fig.\,1(a)), and the shift is proportional to the applied magnetic field. It can also be shown that for lattices having inversion symmetry $\omega_{\sigma}(-\bf k)= \omega_{\sigma}(\bf k)=-\omega_{-\sigma}(\bf k)$ even for nonzero magnetic field. In Fig.\,1(b) we have given a plot both for positive and negative eigenfrequencies $\omega_{\sigma}({\bf k})$ as a function $h$ and we can see that all eigenvalues are real even for very large value of $h$.
\begin{figure}
\includegraphics[width=\columnwidth]{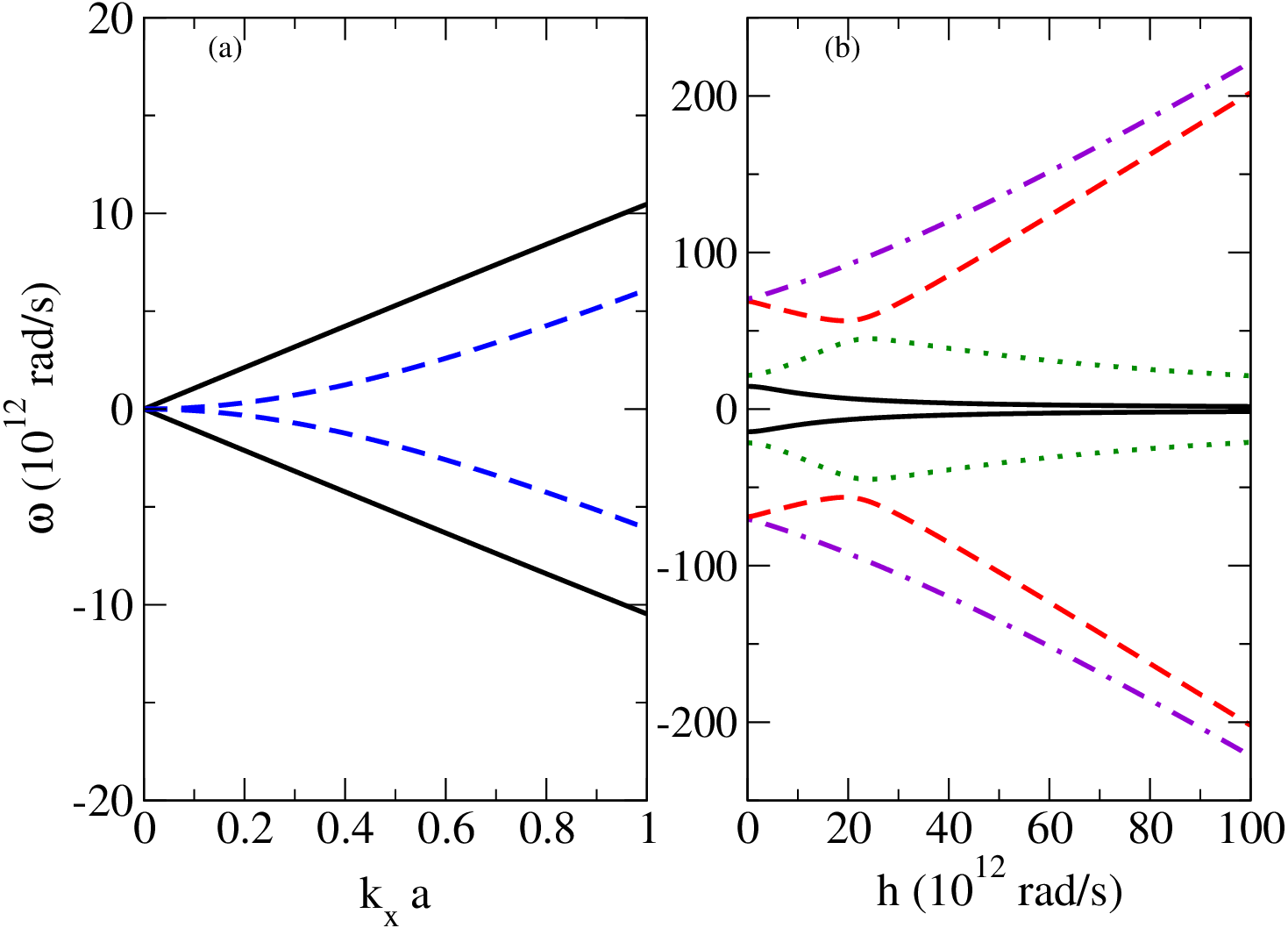}%
\caption{\label{fig1}(Color online)(a) Phonon dispersion curve for hexagonal lattice. The angular frequency (both positive and negative) of one accoustic mode as a function of $k_{x}a$ with $k_{y}=0$. The solid curves are for $h=0$ and dotted curve are for $h=10.0 \times 10^{12}$ rad/s. (b) The eigenfrequencies (two accoustic and two optical) as a function of $h$ for fixed wave vector ${\bf k}a=(1,1)$. }
\end{figure}

\begin{figure}
\begin{tabular}{cc}
\includegraphics[width=\columnwidth]{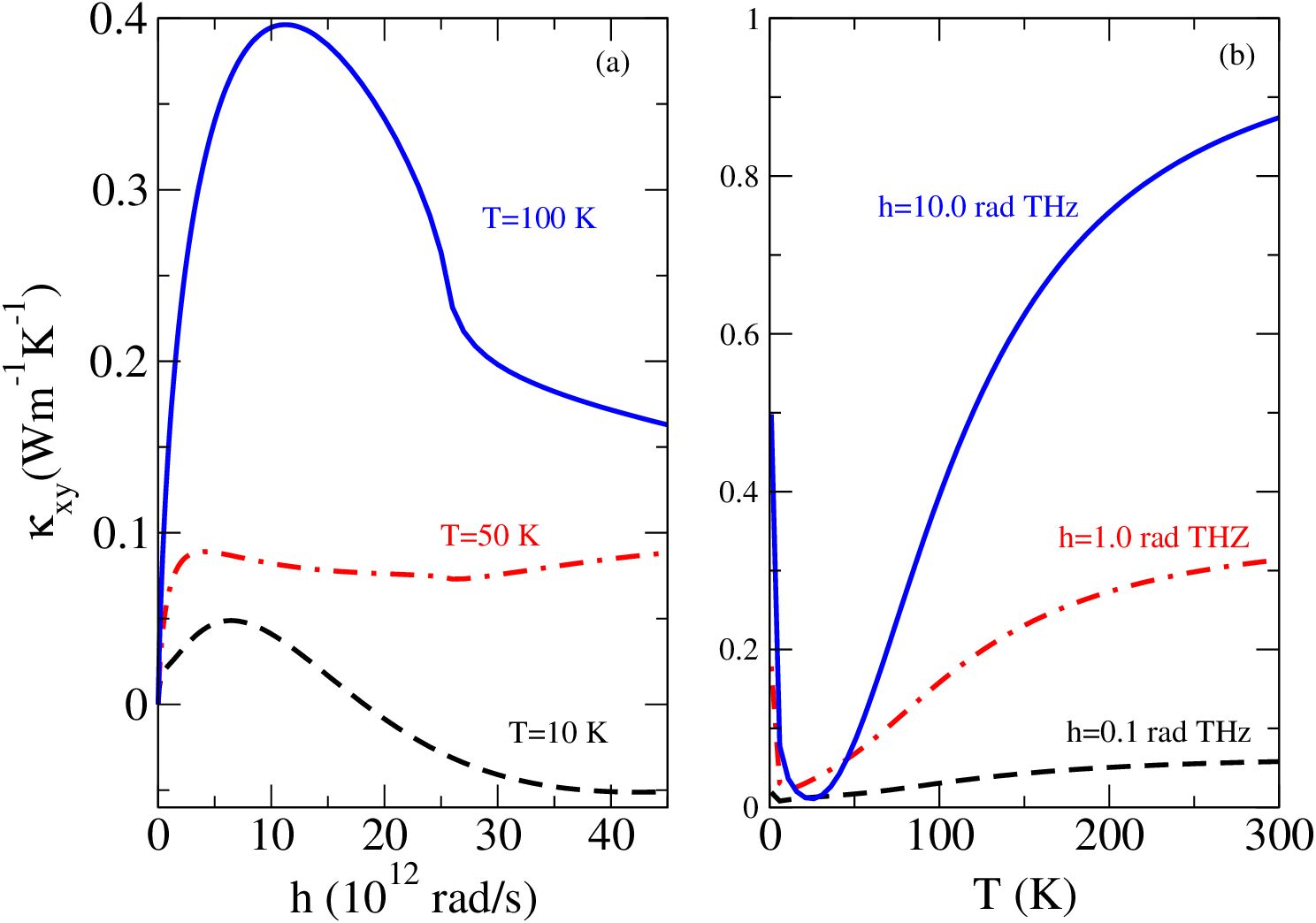}\\
\includegraphics[width=0.8\columnwidth,height=1.9in]{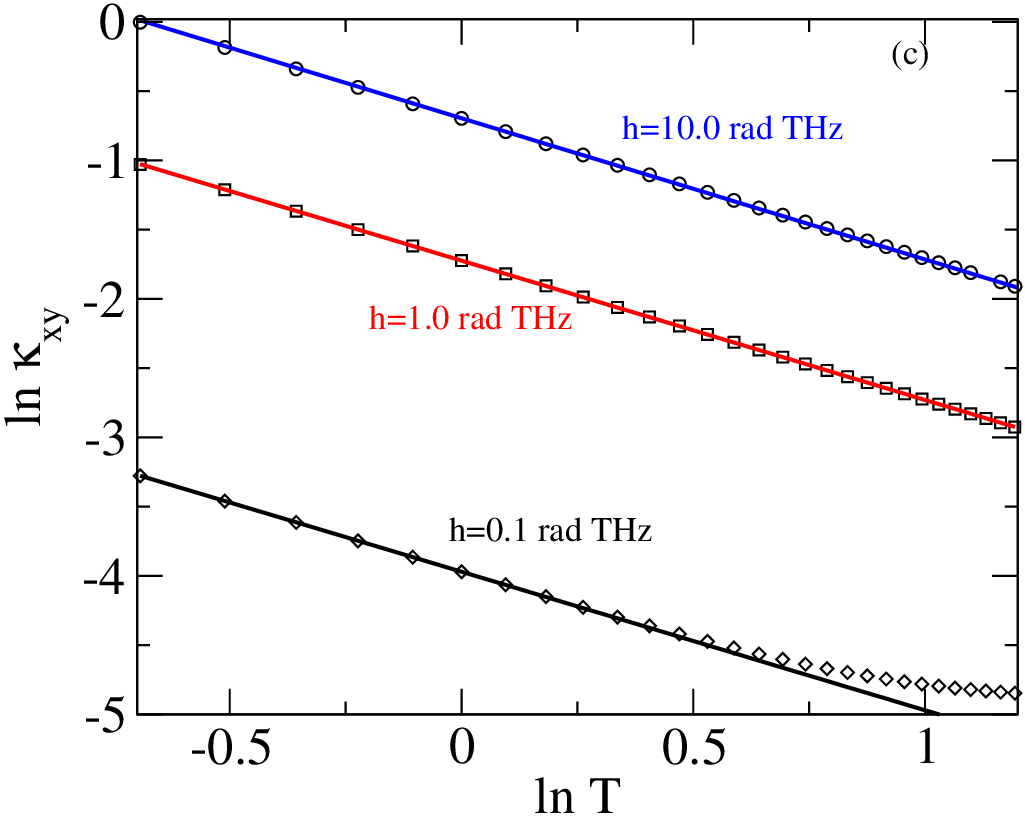}
\end{tabular}
\caption{\label{fig2}(Color online)Thermal Hall conductivity $\kappa_{xy}$ (a) as a function of the
magnetic field $h$ (b) as a function of temperature $T$ (c) divergence of Hall conductivity at low temperature for a hexagonal lattice.}
\end{figure}

In Fig.\,2(a), we show result for the off-diagonal thermal conductivity $\kappa_{xy}$ as a function of magnetic field for three different temperatures $T$=10\,K, 50\,K, and 100\,K. It is clear that, at low temperature for small values of $h$, $\kappa_{xy}$ increases linerly with the magnetic field but for large $h$, it grows slower than linear. For very large $h$ it can even goes to negative value. At low temperature phonon Hall conductivity can change sign with increasing magnetic field. This result is consistent with Ref.~\onlinecite{lifa09}.  With  increasing temperature, the magnitude of negative Hall conductivity decreases because when the temperature is high, more high energy modes are excited and spin-phonon interaction can not easily turn around the phonons.

In Fig.\,2(b) we plot the temperature dependence of $\kappa_{xy}$ for three different values of $h$. For small values of $h$ phonon Hall effect is almost negligable. But for high value of $h$ it increases slowler than linear with $T$ and for very high temperature it reaches a constant value. For low temperature  $\kappa_{xy}^{-1}\sim T^{1.009,1.005,1.012}$ for $h=(10.0,1.0,0.1) \times 10^{12}$ rad/s respectively(see Fig.\,2(c)) which is close to 1.0 and  consistent with Eq.(17). The divergence of $\kappa_{xy}$ as $1/T$ for $T\rightarrow 0$ is not unphysical  as it shows the ballistic behaviour for which the longitudinal conductivity $\kappa_{xx}$ is always infinite at any temperature. However if we measure current it will be zero at $T\rightarrow 0$.  And from another point of view, when $T\rightarrow 0$, the value of $\kappa_{xy}T$ is constant; because of $\Delta T<T$ in linear response region, the measurable quantity of transverse heat current $S^y=-\kappa_{yx} \Delta T/ L$ ( we know $\kappa_{yx}=-\kappa_{xy}$ ) will not be divergent and tend to zero for the periodic system $L \rightarrow \infty$. Furthermore we know that for the ballistic thermal transport, because of the zero temperature gradient in the system, the thermal conductivity is infinite while the thermal conductance is finite and meaningful. Therefore, the $1/T$ divergence of $\kappa_{xy}$ indeed embodies the ballistic picture of PHE at very low temperature. 

\begin{figure}
\includegraphics[width=\columnwidth]{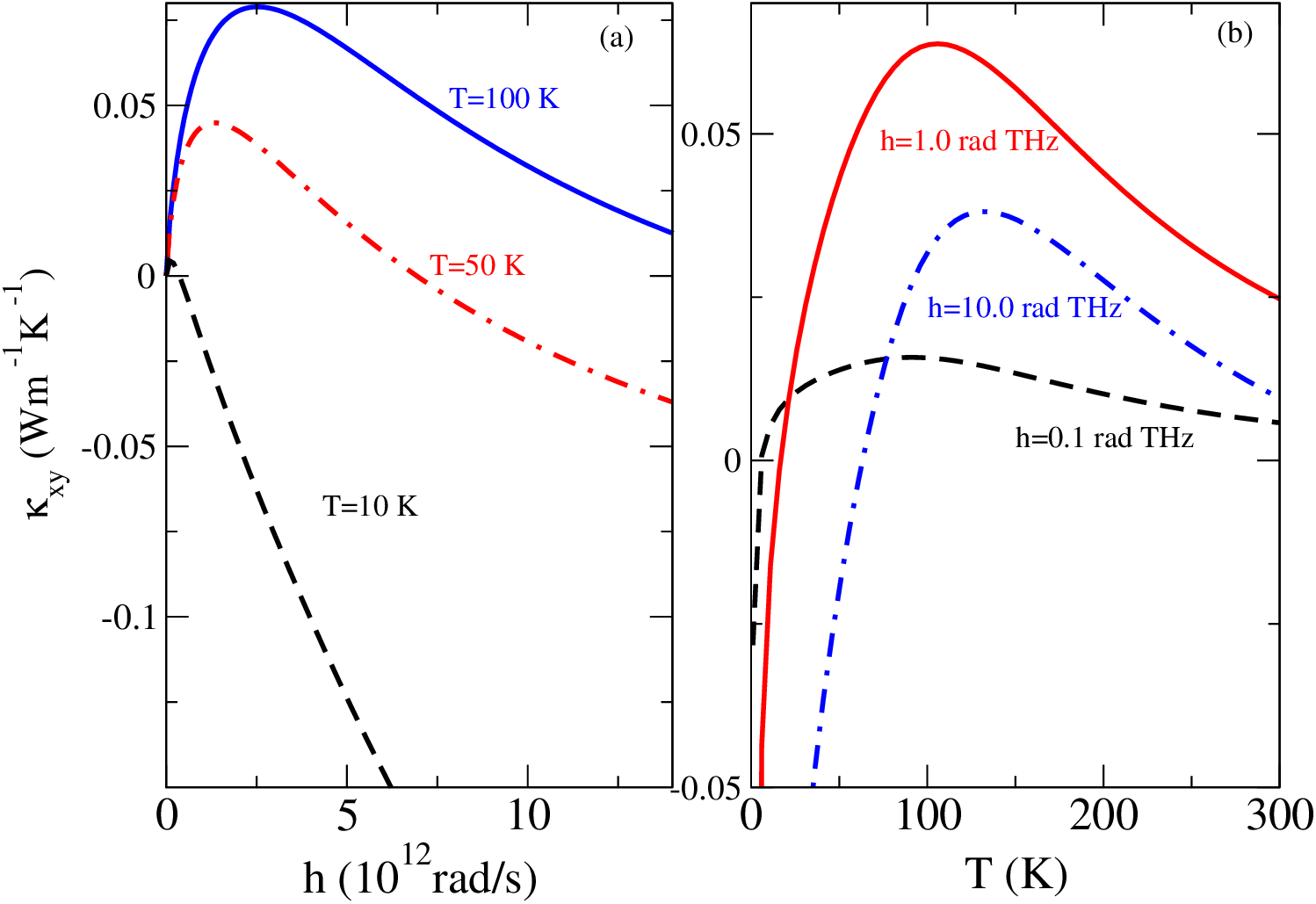}%
\caption{\label{fig3}(Color online)Thermal Hall conductivity $\kappa_{xy}$ (a) as a function of the
magnetic field $h$  and (b) as a function of temperature $T$ for a triangular lattice.}
\end{figure}

\begin{table*}[t]
\caption{Parameters of the potential of NaCl}
\label{tab_nacl}
\begin{ruledtabular}
\begin{tabular}{|l|l|l|l|}
Source & $A$ ($ev$)& $B$(\AA)& $C$ ($ev$)\AA$^{6}$\\
\hline
Na-Na & 487 & 0.23768 & 1.05 \\
\hline
Na-Cl & 14514 & 0.23768 & 6.99  \\
\hline
Cl-Cl & 405774 & 0.23768 & 72.40
\end{tabular}
\end{ruledtabular}
\end{table*}


In Fig 3. we show results for a triangular lattice. We have taken the same values of the parameters for the calculation. In this case also the off-diagonal thermal conductivity shows similar behaviour as in hexagonal lattice. At very low temperature thermal conductivity changes sign. The temperature depedence is also similar. At low temperature conductivity goes as $-1/T$.

For a square lattice with only the nearest neighbour coupling there is no PHE. Generally the off-diagonal term of $\kappa$ will be zero if the reflection symmetry is not broken. If there exists a symmetry operation $S$ such that $S K S^{-1}=K $ and $S A S^{-1}=-A$ then $\kappa_{ab}=0$ for $a\ne b$.  In this case the spring constant matrix $K$ is diagonal and is invariant under reflection in $x$ or $y$ direction, hence it satisfies the above relation. If we consider next-neighbour coupling then the matrix $K$ will not have mirror reflection symmetry and the PHE can be observed which is consistent with Ref.~ \onlinecite{lifa09}.

\section{THE PHE ON THE SODIUM CHLORIDE CRYSTAL}
After the properties of the PHE are well studied for the model lattices, we turn to the calculation for the real ionic crystal NaCl. NaCl crystal can be described as fcc lattice with primitive vectors ${\bf a_{1}}=\frac{a}{2}({\bf y}+{\bf z})$, ${\bf a_{2}}=\frac{a}{2}({\bf x}+{\bf z})$ and ${\bf a_{3}}=\frac{a}{2}({\bf x}+{\bf y})$ and two point basis consisting of a sodium ion at ${\bf 0}$ and a chlorine ion at the center of the conventional cubic cell, $\frac{a}{2}({\bf x}+{\bf y}+{\bf z})$. Here $a$ is the lattice constant which is 5.63 {\AA} for NaCl. 

We calculate the force constant matrix using ``General Utility Lattice Program'' (GULP)\cite{gale97} by considering a unit cell. The interatomic potential we used includes both Coulombic and non-Coulombic terms
\begin{equation}
V_{ij}(r_{i},r_{j})=\frac{1}{4 \pi \epsilon_0} \frac{Z_i Z_j e^{2}}{r_{ij}} + A_{ij}exp(-r_{ij}/B_{ij})-\frac{C_{ij}}{r_{ij}^{6}},
\end{equation}
where $r_{ij}$ is the interatomic distance between ions $i$ and $j$ and $Z_i$ is the effective charge of the $i$th ion. The non-Coulombic part of the potential is known as Buckingham potential. $A_{ij}$ and $B_{ij}$ are parameters for the repulsive interaction and $C_{ij}$ is van der Waals constant. The values of the parameters are given in Table 1.\cite{chin}. Cutoff distance 12 {\AA} is used in the calculation. With this potential model we went upto the third nearest neighbours and cut the force constant matrix properly in order to get the dynamical matrix. The choice of this type of unit cell gives us  dynamical matrix of size $6 \times 6$.

\begin{figure}
\includegraphics[width=\columnwidth]{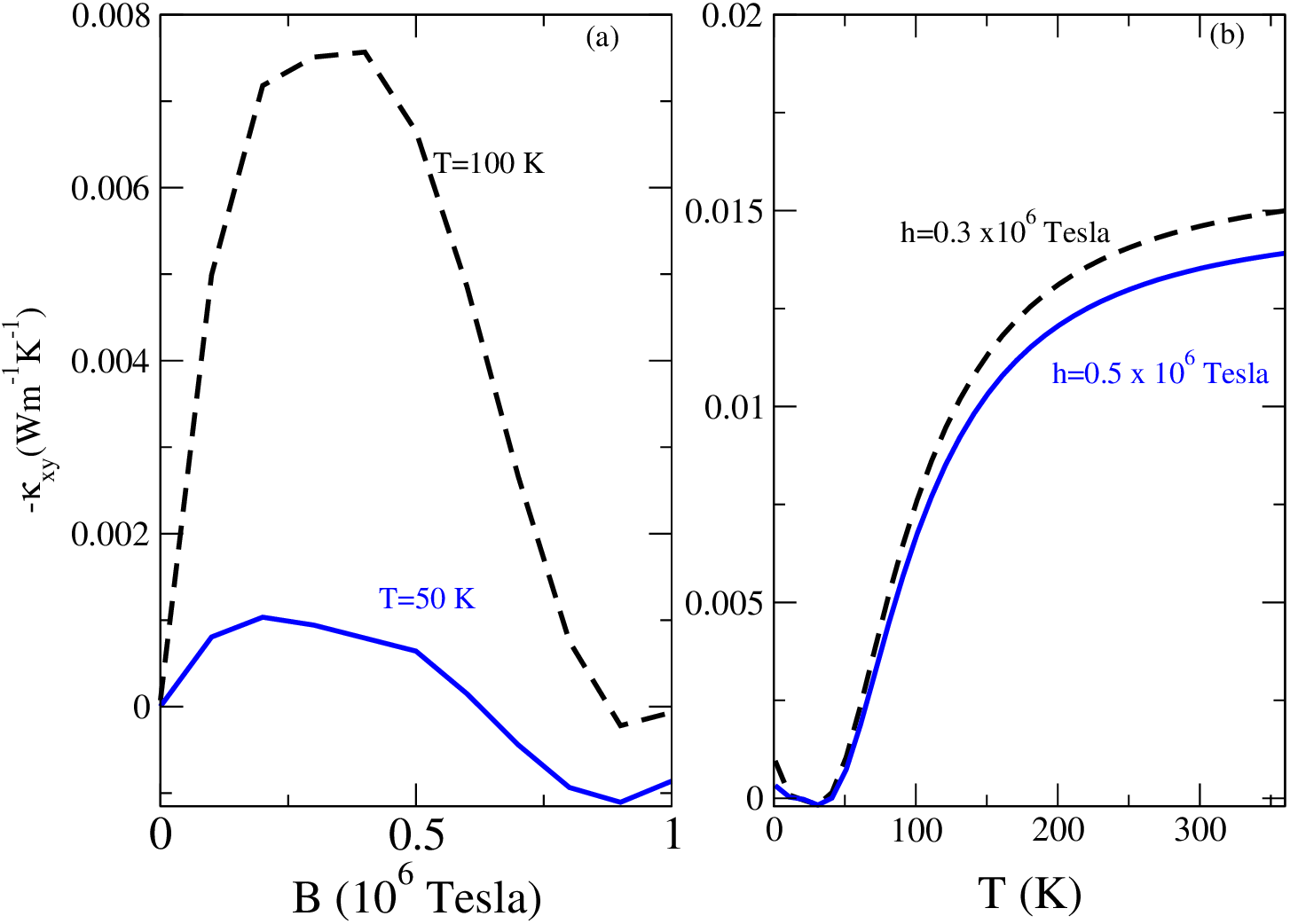}%
\caption{\label{fig4}(Color online)Thermal Hall conductivity $\kappa_{xy}$ (a) as a function of the
magnetic field $B$ (b) and as a function of temperature $T$ for NaCl.}
\end{figure}

In Fig.\,4(a), we give the off-diagonal thermal conductivity $\kappa_{xy}$ as a function of $B$ for temperature $T=100\,K$ and $T=50\,K$. We note that the sign of $\kappa_{xy}$ is negative. The growth of $\kappa_{xy}$ in this case, for small values of magnetic field is very fast, then it reaches a maximum and then decreases to zero with increasing magnetic field and even goes to positive value for high value of magnetic field.


In Fig.\,4(b), we display the temperature dependence of the off-diagonal thermal conductivity $\kappa_{xy}$ for $B=0.3$ and $B=0.5$ MTesla. Upto $T$=40\,K the off diagonal thermal conductivity is close to zero, meaning there is no heat flow. When $T \rightarrow 0$ it also diverge like $1/T$. But it gradually increases and finally saturates at high temperature. In all our calculations we have used $\eta$ of order 10$^{-6}K_{L}$ to get rid of the sigularity due to the accoustic phonons at $\Gamma$ point.

From the above calculation in the nonmagnetic ionic crystal NaCl, we observe the PHE although there is no magnetization in the system. The existence of PHE is due to the nonzero term of $\tilde{A}$ in Eq.~(\ref{eq-ham}). Both the positive and negative ions in the crystal contribute to $\tilde{A}$, and they can not calcel each other. Since in the term $\tilde{A}$, the charge to mass ratio $ (e/m)$ is very small for nonmagnetic ionic crystals NaCl, a very strong magnetic field is required to observe PHE in this case. However in the paramagnetic crystal the contribution from electrons dominates the term ${\tilde A}$, which increases about $10^{4}$ times and thus the PHE can be comparable to the experimental data. For ferromagnetic insulating ionic crystal, it will be increased much more because of the large magnetization which will contribute to large spin-phonon interaction.

\section{CONCLUSION}
A theory for phonon Hall effect in general ionic crystals based on a ballistic lattice dynamic model is developed. An exact formula of phonon Hall conductivity formula from Green-Kubo approach is derived. It is shown that the effect can be present in the presence of a very strong magnetic field for general ionic crystals without magnetization. Due to the tiny charge-to-mass ratio, the PHE is so small that it is difficult to be measured experimentally. However the theoretically confirmed existence of PHE can give us deep understanding of the PHE and guide us the direction to do the experiments. The ferromagnetic insulator is a good candidate for the PHE. We hope some artificial lattices such as cool atoms in optical trap can generate similar kind of effect. We believe that the result for $\kappa_{xy}$ is robust against nonlinear perturbation, but it is interesting to work out a theory where nonlinearity is taken into account.

\section{ACKNOWLEDGEMENTS}
we are greatful to Mr. Juzar Thingna, Dr. Jin-Wu Jiang and Meng Lee Leek for insightful discussions. This work is supported in part by URC grant R-144-000-257-112 and R-144-000-222-646 of National University of Singapore.

\appendix
\section{Second Quantization}
For the negative phonon branches we define the polarization vectors as following
\begin{eqnarray}
 \epsilon_{-\sigma}(-{\bf k})&=&\epsilon_{\sigma}^{\ast}({\bf k}),\\
\omega_{-\sigma}(-{\bf k}) &=& -\omega_{\sigma}({\bf k}).
\end{eqnarray}
This defination is consistent with Eq. (7). Also the creation and annihilation operators are defined as for $\sigma>0$ :
\begin{equation}
a_{-\sigma}({\bf k})=a_{\sigma}^{\dagger}(-{\bf k}),
\end{equation}
and the time dependence is given by for $\sigma>0$ :
\begin{eqnarray}
a_{\sigma}({\bf k})(t)&=& a_{\sigma}({\bf k})e^{-i \omega_{\sigma}({\bf k})t},\\
a_{\sigma}^{\dagger}({\bf k})(t)&=& a_{\sigma}^{\dagger}({\bf k})e^{i \omega_{\sigma}({\bf k})t}.
\end{eqnarray}
The commutaion relation between the operators is
\begin{equation}
[a_{\sigma}({\bf k}), a_{\sigma'}({\bf k'})]=\delta{{\bf (k+k')}} \delta{(\sigma+\sigma')}  \rm{sign}(\sigma).
\end{equation}

The matrix $\cal M$ in Eq.(8) is not a normal matrix. Hence we need to consider both the left and right eigenvectors. However it can be easily shown that they are not really independent. For a given value of $\omega_{\sigma}({\bf k})$ and corresponding right eigenvector $\chi_{\sigma}({\bf k})=(\mu_{\sigma}({\bf k}), \epsilon_{\sigma}({\bf k}))^T$ one can choose the left eigenvector as  $\tilde \chi_{\sigma}^T({\bf k})=(\epsilon_{\sigma}^{\dagger}({\bf k})$/$(-2 i \omega_{\sigma}({\bf k})), -\mu_{\sigma}^{\dagger}({\bf k})$/$(-2 i \omega_{\sigma}({\bf k})))$. With this choice of eigenvectors the normalization can be done according to
\begin{equation}
\epsilon_{\sigma}^\dagger({\bf k})\cdot \epsilon_{\sigma}({\bf k}) + \frac{i}{\omega_{\sigma}({\bf k})} \epsilon_{\sigma}^\dagger({\bf k})\cdot {\tilde A} \cdot \epsilon_{\sigma}({\bf k)} = 1.
\end{equation}
The Hamiltonian in Eq. (3) can be written as
\begin{equation}
H=\frac{1}{2}\sum_{l,l'} \tilde{\mathbb{X}}^T_{l} \left( \begin{array}{cc} {\tilde A} \delta_{l,l'} & K_{l,l'} \\
                      -I \delta_{l,l'} & {\tilde A} \delta_{l,l'} \end{array} \right) {\mathbb{X}}_{l'}
\end{equation}
where
\begin{eqnarray}
{\mathbb{X}}_{l}=\left( \begin {array}{rr}p_{l} \\ u_{l} \end{array} \right)&=&\sum_{{\bf k},\sigma} \chi_{\sigma}({\bf k}) e^{i {\bf R}_l \cdot {\bf k}} \sqrt{\frac{\hbar}{2\left|\omega_{\sigma}({\bf k})\right | N}}\; a_{\sigma}({\bf k}); \nonumber \\
\tilde{\mathbb{X}}_{l} = \left( \begin {array}{rr}u_{l} \\ -p_{l} \end{array} \right)&=&\sum_{{\bf k},\sigma}(-2i\omega_{\sigma}({\bf k})) \tilde \chi_{\sigma}({\bf k})e^{-i {\bf R}_l \cdot {\bf k}}\times\nonumber \\
&&\sqrt{\frac{\hbar}{2\left|\omega_{\sigma}(\bf k)\right| N}} a_{\sigma}^{\dagger}({\bf k}),
\end{eqnarray}
where we have used the definition for $u_{l}$ and $p_{l}$ and also the Hermitian property of the operator.
Now using the definition of dynamical matrix, the identity that $\sum_{l}e^{i({\bf k'}-{\bf k}) \cdot {\bf R}_l}=N\delta({\bf k'}-{\bf k})$, and the orthogonality relation between left and right eigenvector $\tilde \chi_{\sigma}^T({\bf k})\cdot \chi_{\sigma'}({\bf k})=\delta_{\sigma \sigma'}$, the Hamiltonian reduces to the form of Eq. (11). The canonical commutation relation can be shown to be satisfied using the completeness of $\chi$, as
\begin{equation}
[{\mathbb{X}}_{l},\tilde{\mathbb{X}}^T_{l'}]=-i\hbar \delta_{ll'}I.
\end{equation}

\section{Positive definiteness of the Hamiltonian}
To prove that the Hamiltonian is positive definite we can write the Hamiltonian as
\begin{equation}
H=\frac{1}{2} \left( \begin {array}{ll}u & p \end{array} \right) \left( \begin{array}{cc} K & 2{\tilde A} \\
                      0 & I \end{array} \right)  \left( \begin {array}{rr}u \\ p \end{array} \right),
\label{eq-positive}
\end{equation}
with $K=\phi -{\tilde A}^{2}$. We assume that $\phi$ is positive definite.
Since transpose of a positive definite matrix is positive definite we can as well consider the matrix
\begin{equation}
\cal Q=\left( \begin{array}{cc} K & {\tilde A} \\
                      -{\tilde A} & I \end{array} \right) ,
\label{eq-positive1}
\end{equation}
which is a symmetric matrix. Hence it can be diagonalized by similarity transformation and can be written as $\cal Q=S D S^T$
where $\cal D=$diag$(\lambda_1,\lambda_2, ......)$. Now if $\lambda_i \ge 0$ $\forall$ $i$ then $\cal D$$^ \frac{1}{2}$ exists and we can write
$\cal Q=B^T B $ with $\cal B=D$$^ \frac{1}{2} S^T$. So a real symmetric matrix has non-negative eigenvalues iff it can be factorized as $\cal B^T B$.
The above matrix $\cal Q$ can be factorized according to
\begin{equation}
\cal B=\left( \begin{array}{cc} -{\tilde A} & I \\
                      \sqrt{(K+{\tilde A}^{2})^{T}} & 0 \end{array} \right).
\label{eq-positive}
\end{equation}
Hence the Hamiltonian is positive definite.


\end{document}